\documentclass{sigchi}

\newfont{\mycrnotice}{ptmr8t at 7pt}
\newfont{\myconfname}{ptmri8t at 7pt}
\let\crnotice\mycrnotice%
\let\confname\myconfname%
\CopyrightYear{2020}
\setcopyright{acmlicensed}
\doi{https://doi.org/10.1145/3313831.XXXXXXX}
\isbn{978-1-4503-6708-0/20/04}
\conferenceinfo{CHI'20,}{April  25--30, 2020, Honolulu, HI, USA}
\acmPrice{\$15.00}
\usepackage{balance}       
\usepackage{graphics}      
\usepackage[T1]{fontenc}   
\usepackage{txfonts}
\usepackage{mathptmx}
\usepackage{hyperref}
\usepackage{color}
\usepackage{booktabs}
\usepackage{textcomp}

\usepackage{microtype}        
\usepackage{ccicons}          

\usepackage{todonotes}

\usepackage{xac}

\def\plaintitle{CheXplain: Enabling Physicians to Explore and Understand Data-Driven, AI-Enabled Medical Imaging Analysis}

\def\emptyauthor{}
\def\plainkeywords{Authors' choice; of terms; separated; by
  semicolons; include commas, within terms only; this section is required.}

\makeatletter
\def\url@leostyle{%
  \@ifundefined{selectfont}{
    \def\UrlFont{\sf}
  }{
    \def\UrlFont{\small\bf\ttfamily}
  }}
\makeatother
\def\pprw{8.5in}
\def\pprh{11in}

\definecolor{linkColor}{RGB}{6,125,233}
\hypersetup{%
  pdftitle={\plaintitle},
  pdfauthor={\emptyauthor},
  pdfkeywords={\plainkeywords},
  pdfdisplaydoctitle=true, 
  bookmarksnumbered,
  pdfstartview={FitH},
  colorlinks,
  citecolor=black,
  filecolor=black,
  linkcolor=black,
  urlcolor=linkColor,
  breaklinks=true,
  hypertexnames=false
}

\begin{document}

\title{\plaintitle}

\numberofauthors{3}
\author{
   \alignauthor{Yao Xie\\
     \affaddr{HCI Research, UCLA}\\
     \affaddr{Los Angeles, California, United States}\\
     \email{yaoxie@ucla.edu}}\\
  \alignauthor{Melody Chen\\
     \affaddr{HCI Research, UCLA}\\
     \affaddr{Los Angeles, California, United States}\\
     \email{melodyc1205@ucla.edu}}\\
  \alignauthor{David Kao\\
     \affaddr{HCI Research, UCLA}\\
     \affaddr{Los Angeles, California, United States}\\
     \email{davidkao41@ucla.edu}}\\
  \alignauthor{Ge Gao\\
     \affaddr{College of Information Studies, University of Maryland, College Park}\\
     \affaddr{Maryland, United States}\\
     \email{gegao@umd.edu}}\\
   \alignauthor{Xiang 'Anthony' Chen \\
     \affaddr{HCI Research, UCLA}\\
     \affaddr{Los Angeles, California, United States}\\
     \email{xac@ucla.edu}}\\
}

\maketitle

\begin{abstract}
  The recent development of data-driven AI promises to automate medical diagnosis; however, most AI functions as `black boxes' to physicians with limited computational knowledge. Using medical imaging as a point of departure, we conducted three iterations of design activities to formulate CheXplain---a system that enables physicians to explore and understand AI-enabled chest X-ray analysis: \one a paired survey between referring physicians and radiologists reveals whether, when, and what kinds of explanations are needed; \two a low-fidelity prototype co-designed with three physicians formulates eight key features; and \three a high-fidelity prototype evaluated by another six physicians provides detailed summative insights on how each feature enables the exploration and understanding of AI. We summarize by discussing recommendations for future work to design and implement explainable medical AI systems that encompass four recurring themes: motivation, constraint, explanation, and justification.
\end{abstract}




\section{Introduction}
\begin{figure}
  \includegraphics[width=\linewidth]{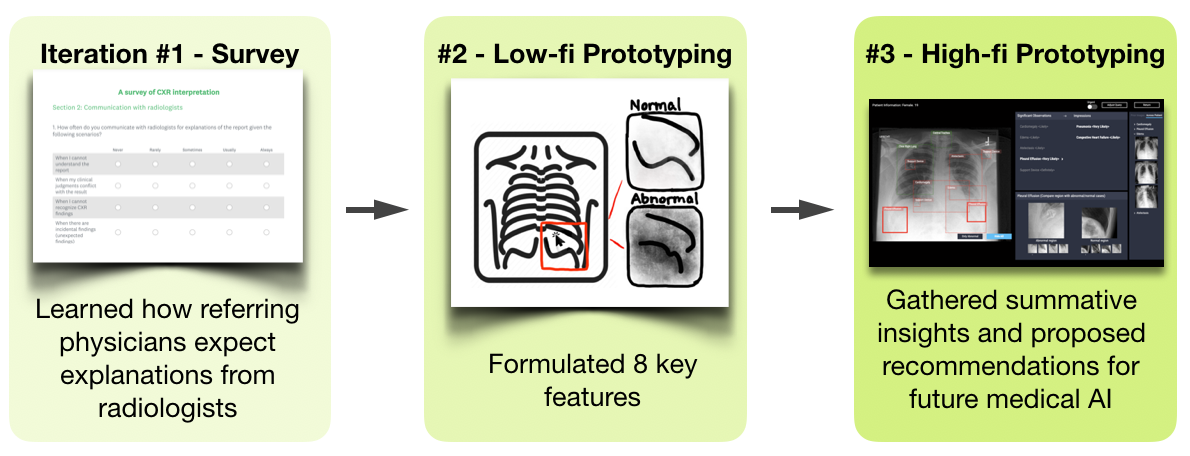}
  \caption{We conducted three user-centered activities to formulate the design of CheXplain --- a system that enables physicians to explore and understand AI-enabled chest X-ray analysis. \one A paired survey study learns about how referring physicians expect explanations from radiologists. \two A user-centered design with three physicians formulates eight design ideas as low-fi prototypes. \three An integrated high-fi prototype evaluated by six physicians provides insights and recommendations for future development of explainable medical AI systems.}

    \label{fg:fig1}
    \vspace{-2em}
\end{figure}
Intelligent systems---computational agents that employ artificial intelligence (AI) to process and make sense of data---are becoming increasingly ubiquitous in modern workplaces \cite{Abdul2018-nz}. For example, stakeholders use algorithms to assist them in urban planning, predicting the risk of future crimes or estimating insurance risk \cite{Thebault-Spieker2017-pa,Zhao2017-ug}. Despite the promise of assisting human decision making through a data-driven approach, non-computing professionals often find it challenging to understand how a ‘black box’ system transforms their initial input into a final decision.

In medicine, the `black box' problem of AI also raised concerns. Started in the 1950s \cite{Miller2018-qs}, medical AI has been widely applied in many subfields such as early detection, diagnosis, and treatment planning \cite{Jiang2017-ex}. Early work on medical expert systems such as DXplain \cite{Barnett1987-pv} was able to incorporate rule-based explanation as part of the automatically generated results.

The recent development of deep neural networks promises that data-driven AI can process diagnostic imaging on par with human physicians \cite{Irvin2019-hc}. However, compared with expert systems, neural networks rely on very low-level constructs---a large number of neurons distributed across stacks of layers, making it challenging, if not impossible, for medical professionals to understand how a data-driven AI arrives at certain conclusions. Such a lack of transparency creates a barrier of understanding, preventing AI from being widely adopted in the clinic despite of its promising performance \cite{Zheng2019-mu}.

To solve the `black box' problem, prior work has been focusing on developing interpretable, accountable and transparent algorithms \cite{Diakopoulos2016-cd,Shneiderman2016-en}, visualizing obscure features in medical images \cite{Donahue2014-tp,Zeiler2014-rr} or employing cognitive psychological theories to explore effective explanations \cite{Lombrozo2010-yq,Lombrozo2006-uk,Miller2018-qs}. Little of this work, however, approaches AI's explainability problem from the physicians' point of view, \ie considering physicians' domain-specific needs and day-to-day practices.

Our goal is to enable physicians to understand existing data-driven, AI-enabled medical imaging analysis. We choose imaging as it is the primary data source in medicine \cite{Jiang2017-ex}. In this paper, we focus on one of the most common modalities---chest X-ray (CXR) images. To achieve this goal, as shown in \fgref{fig1}, we started with a survey, which informed a user-centered design to formulate CheXplain---a system for physicians to interactively explore and understand CXR analysis generated by a state-of-the-art AI \cite{Irvin2019-hc}.

    {\bf Iteration \#1: Survey} We conducted a paired-survey on both referring physicians (N=39) and radiologists (N=38) to learn about how radiologists currently explain their analysis to referring physicians and how referring physicians expect explanations from both human and (hypothetical) AI radiologists. The findings reveal whether, when, and what kinds of explanations are needed between referring physicians and radiologists. By juxtaposing referring physicians' response to these questions with respect to human vs. AI radiologists, we elicit system requirements that encompass the current practices and a future of AI-enabled diagnosis.

    {\bf Iteration \#2: Low-fi prototype}. A co-design with three physicians manifested the survey-generated system requirements into eight specific features: augmenting input CXR images with specific inquiries, mediating system complexity by the level of urgency, presenting explanations hierarchically, calibrating regional findings with contrastive examples, communicating results probabilistically, contextualizing impressions with additional information, and comparing with images in the past or from other patients.

    {\bf Iteration \#3: High-fi prototype}. We integrate the eight features into CheXplain---a functional system front-end that allows physician users to explore {\it real} results generated by an AI \cite{Irvin2019-hc} while all the other explanatory information either comes from real clinical reports or is manually generated by a radiologist collaborator. An evaluation with six medical professionals provides summative insights on each feature. Participants provided more detailed and accurate explanations of the underlying AI after interacting with CheXplain. We summarize the design and implementation recommendations for future development of explainable medical AI.

\subsection{Contributions}
Our contributions are as follows:
\begin{itemize}
    \item An empirically-informed, user-centered iterative design of CheXplain---a proof-of-concept system prototype that integrates eight key features for enabling physicians to explore and understand AI-generated medical imaging analysis;
    \item A summary of design and implementation recommendations for future development of explainable medical AI, encompassing four recurring themes: motivation, constraints, explanation, and justification. For example:
    
    \begin{itemize}
        \item Insight: often what referring physicians look for is not explanation, but \textit{justification}---information extrinsic to a CXR image that can support an AI's analysis result.
        \item Recommendation: more explanation for medical data analysis (\eg radiologists), and more justification for medical data consumers (\eg referring physicians).
    \end{itemize}

\end{itemize}


\section{Background: AI and Radiology}

The advent of data-driven AI has given rise to a plethora of research and development using AI to perform radiographic analysis \cite{Miller2019-sd,shin2016learning,takemiya2019detection}. The recent work on CheXpert \cite{Irvin2019-hc} developed a deep neural network with a dataset that contains 224,316 chest X-ray images from 65,240 patients with 14 different observations. Such an `AI radiologist' takes single CXR image as the input and outputs the labels that consist of a list of observations. The model in CheXpert has achieved a radiologist-level performance---an accuracy of 83\% based on ground truths from CXR reports---and importantly, much faster diagnosis than human radiologists.
\subsection{Nomenclatures}
\begin{itemize}
    \item Referring physician: a non-radiologist physician who sends imaging data (\eg CXR images) of a patient to a specialist for more information or treatment.
    \item Observation: in radiology reports, an observation is a description or finding from the radiograph.
    \item Impression: in radiology reports, impression comes as a list of summary statements or differential diagnoses concluded from the observations.

\end{itemize}

\section{Related Work}
We review explainable AI (XAI) and its recent development in the HCI community. We then introduce literature in Clinical Decision Support Systems (CDSS), and since our focus is on imaging, we further discuss prior work on the communication between radiologists and referring physicians.

\subsection{Explainable Artificial Intelligence (XAI) Systems}
Researchers have focused on the explanations of expert intelligent systems since the 1970's \cite{Van_Melle1984-lm}. The topic of explainability was rejuvaneted as recent developments in data-driven AI (\eg deep learning) exacerbates the obscurity of the hidden intelligent process. Ras \etal   summarized recent XAI research into four topics: users, laws/regulations, explanations and algorithms \cite{Ras2018-ry}. A plethora of work on interpretable machine learning has sought to explain the inner principles of the machine learning models with mathematical and algorithmic solutions \cite{Biran2017-dx}. However, there is no clear agreement on the definition of interpretability and controversy still exists. Lipton \cite{Lipton2016-fi} proposed the motivations and methods to achieve interpretability. Doshi-Velez and Kim \cite{Doshi-Velez2017-vj} discussed how to measure interpretability. Miller \cite{Miller2019-sd} defined interpretability as the degree to which a human can understand the cause of a decision and provided an overview of research about how people define, generate, validate, on present explanations.

To explain data-driven AI, researchers have focused on three categories of XAI solutions: rule-extraction methods, attribution methods, and intrinsic methods \cite{Ras2018-ry}. 

\begin{itemize}
    \item Rule-extraction methods generate rules based on input and output information to simulate the ‘black box' decision-making process. For example, Frosst and Hinton distilled a more understandable soft decision tree from a neural network \cite{Frosst2017-vs}. Murdoch and Szlam constructed a simple, rule-based classifier that approximates the output of LSTM \cite{James_Murdoch2017-nw}. Ribeiro \etal developed LIME, which learns the model locally around the prediction and explains the prediction in an interpretable and faithful manner \cite{Ribeiro2016-vc}. 
    \item Attribution methods measure the importance of a component to the output by changing the input or internal components. DeepLIFT decomposed the output prediction of a specific input by backpropagating the contribution of all the neurons to the input features \cite{Shrikumar2017-lq}. Zintgraf \etal proposed a method of visualizing the response of a DNN to a specific input to analyze the prediction difference \cite{Zintgraf2017-ey}.
    \item Intrinsic methods enhance the interpretability of a model's internal constructs without affecting its performance. Santoro \etal transformed a DNN model into one that is able to answer rationale questions by adding a reasoning module that learns a relational function \cite{Santoro2017-nv}. Others used a framework that consists of an attentive encoder-decoder network for video caption generation and an interpretive loss to visualize the learned features \cite{Dong2017-zk, Wu2017-tl}.
\end{itemize}

Besides the works in ML and AI communities, in the HCI community, people attempt to understand AI from the users' perspective, which we review below.

\subsection{Explainable Artificial Intelligence (XAI) Research in HCI}
Intelligent sensing systems need to provide users with not only results but also accounts for their behaviors \cite{Bellotti2001-pm}. XAI in HCI contains topics that intersect context awareness, cognitive psychology, and software learnability \cite{Shneiderman2016-en}. 
For explainable context-awareness, the key is using simplistic representations of the context to inform users what is obtained and which action will be done by the systems \cite{Dourish1995-ox}. Dey and Newberger designed a tailored interface that provides visual and textual explanations for context-aware rules \cite{Dey2009-fw}. Cognitive psychology focuses more on explanation theory. Lombrozo studied cognitive explanations and found that it is strongly connected with causality reasoning \cite{Lombrozo2010-yq}. Software learnability is an important part of usability \cite{Grossman2010-yu} and Abdul summarized topics of learnability such as hints, guidance as well as visualizations that are related to design an XAI system \cite{Abdul2018-nz}. Another study proposed a framework emphasizing empirical application-specific investigations of XAI by exploring the theoretical underpinnings of human decision-making \cite{Wang2019-if}. Others focused on information visualization and ML algorithms visualization to help non-expert users understand AI \cite{Bussone2015-te}.


However, the research in the HCI and AI communities often seem to be disconnected \cite{Abdul2018-nz}. There is a lack of research that crosses and combines both fields to interdisciplinarily approach the XAI problem.

\subsection{Clinical Decision Support Systems (CDSS)}
Clinical Decision Support Systems (CDSS) provide knowledge to enhance medical decision-making for physicians. CDSS is widely adopted in fields such as screening and prevention, medication decision-making, therapeutic planning, and diagnostics \cite{Bussone2015-te}. 
With the advent of available medical data and data processing techniques, CDSS have become increasingly powerful in extracting useful information from a large amount of patient data to assist medical decision making \cite{Fieschi2013-re,Murdoch2013-ni}. AI-based Medical Diagnosis Support Systems (AIMDSS) have become an important part of CDSS. Currently, research in AIMDSS primarily focuses on pathology and radiology, with the ability to detect nuanced patterns that are otherwise hardly noticeable by human doctors. For example, Jha and Topol mentioned AIMDSS is able to identify radiographs for radiologists and recognize patterns for pathologists to work as a information specialist during the diagnostic process \cite{Jha2016-wf}. 

However, the actual usage of these AI-based Clinical Decision Support Systems is limited because of a lack of understanding. Fan \etal~found that initial  and performance expectancy both have significant effects on doctors' behavioral intention of using AIMDSS \cite{Fan2018-fh}. 
Medical professionals may not use a system if the system cannot provide relevant information or capture the mental model of doctors \cite{Khairat2018-ks,Kohli2018-tc, Yang2016-vz}. Thus it is crucial that CDSS are able to explain themselves. The explanation capabilities of AI systems using knowledge bases were first added to medical decision making and computer-aided diagnosis in 1983 \cite{Swartout1983-rv}. MYCIN was one of the earliest systems that incorporated domain knowledge and rule representations to achieve explainability \cite{Van_Melle1984-lm}. DXplain explained and justified the interpretations by proposing knowledge-based diagnostic hypotheses including signs and symptoms to users in medical decision-making \cite{Barnett1987-pv}. A diagnostic reasoning theory is proposed to find the components of the system that explain the discrepancy between the expected result and observed behaviors \cite{Reiter1987-ge}. 
%
An argumentation-based interaction \cite{Fox2007-ad} that is flexible and easily understood by human users can help doctors make decisions based on this question. Better explanations can let users better understand the reasoning chain and enhance the system's confidence \cite{Bussone2015-te}.  Park and Han proposed explanations as one of the methods of evaluating the clinical performance and effect of CDSS \cite{Park2018-kf}. 
%

New research also emerged to address explainability challenges in medical AI systems that started to incorporate data-driven models.
Caruana \etal developed an interpretable system using the generalized additive models with pairwise interactions $GA^{2}M$ model predicting pneumonia risk and hospital 30-day readmission \cite{Caruana2015-ub}. Krause \etal designed a visual analytics system to help support interactive dependence diagnostics by feature representation and visualization \cite{Krause2016-aq}.

In sum, research in CDSS attempted to bridge the gap of actual use by making it understandable to domain experts in medical decision-making. As this paper is conerned with XAI in radiology, below we review past research on the exchange of information between radiologists and referring physicians to understand their decision-making process.

\subsection{Communication between Radiologists and Physicians}

One study mentioned that the ambiguous requests from referring physicians may prevent radiologists from assisting physicians efficiently and a more precise medical history should be helpful for radiologists to better understand the cases \cite{Kilhenny1972-hr}. 
Fischer \etal found that direct communication, informative request forms and questioning the patients will also be useful to lead to better communication between referring physicians and radiologists \cite{Fischer1983-pi}. 
Another study proposed specific suggestions on how to improve reporting performance from referring physicians' points of view. The study indicated that reporting should receive more attention in training and practice than it currently does \cite{Gunderman2000-dx}. 
The American College of Radiology (ACR) also defined practice guidelines for communication of diagnostic imaging findings for radiologists aiming for better communication \cite{Acr2010-xa}. 
In addition, the quality of requests, transparency with regard to waiting time, a portal for easier scheduling, and making radiation dose information visible were also mentioned in a survey \cite{Sectra2013-ii}. 
One study pointed out that the vagueness of the radiology report, \eg inappropriate use of words in radiology reports, would lead to information asymmetry between referring physicians and radiologists, which might prevent timely treatments \cite{Valls2001-xm}. Clinger mentioned referring physicians' satisfactory is mainly affected by diagnostic accuracy, clarity of language, and a detailed discussion of the findings \cite{clinger1988radiology}.
Berlin pointed out that radiologists' duty extends beyond interpreting and sending out a written report to referring physicians. The communication of diagnosis is altogether as important as the diagnosis itself \cite{Berlin2007-vs}.

Despite all this research, little has been studied about how explanation---a specific form of communication---is currently used between referring physicians and radiologists, not to mention what difference there will be in an AI-enabled scenario. To bridge this gap and to inform the design of our system, below we started with a survey of explanations between referring physicians and radiologists.


\section{Iteration \#1: Learning About Explanations\\Between Physicians \& Radiologists}
Given the limited prior art on explanation between physicians and radiologists, we conducted a survey to set the scene for the subsequent design activities. Specifically, we investigated whether, when, and what kinds of explanations are needed between referring physicians and radiologists when performing a diagnosis based on CXR images.

\subsection{Survey Design}
The authors all have computer science backgrounds. To understand communications between radiologists and referring physicians in real workflow, and to design survey questions, we conducted several interviews with two radiologists.

We employed a paired-survey design where both referring physicians and radiologists would answer the same set of questions, respectively, as explanation involves both an explainer and an explainee. For example, physicians would answer ``what kinds of information do radiologists use to explain their examination results to you? '', whereas the same question for the radiologists would be ``what kinds of information do you use to explain your examination results to the referring physicians?''. This design allows us to compare and synthesize both parties' response \textit{vis-à-vis} in order to obtain a more complete picture of explanations between referring physicians and radiologists.

Further, we asked another set of questions only to referring physicians, specifically about when and what kinds of explanations they expect if the results were to come from an AI rather than a human radiologist. Given that very few AI-enabled radiograph analysis tools exist in clinical use, this set of questions is by nature speculative. Based on the state-of-the-art development \cite{Irvin2019-hc}, we described an AI radiologist to participants as a computer program that can produce---in real time---observations commonly found in human radiologists' report. This design allows us to identify if there is any difference in referring physicians' expectations of explanation when it comes to AI.

\subsection{Participants}
We distributed our surveys via social media groups of medical professionals across the US and personal connections to local medical centers. We recruited 77 participants in total (39 referring physicians and 38 radiologists). The study was conducted anonymously on Survey Monkey Platform and each participant received \$20 as payment. \label{survey demographic}


\subsection{Findings}
\subsubsection{Whether Explanation Is Needed}
\textit{In general, a lack of information motivates but the time cost constrains the seeking of explanations}.
Referring physicians often find themsevles in need for more information, \eg ``\textit{more comparisons}'', ``\textit{more details instead of vague descriptions of abnormalities}''. Calling radiologists or colleagues for explanation is considered ``\textit{one effective way to obtain information}''. 
In the meantime, as is similar to the process of reporting \cite{bosmans2011radiology}, seeking explanation is also time-consuming: participants reported that waiting time for initiate a communication ranges from 15 minutes to one day, where only 56\% of the participants felt satisfied with such efficiency.

\textit{With AI, a lack of trust becomes the motivating factor}. Referring physicians remain skeptical, as only $1/3$ of them expressed trust of an AI radiologist, compared to $38/39$ reported likely or very likely to trust a human radiologist. Participants indicated that an AI radiologist would need to explain itself with ``\textit{probabilities}'' and ``\textit{professional knowledge}''.

\subsubsection{When Explanation Is Needed}
\textit{Referring physicians seek explanations when their questions are unanswered or when the results are atypical or critical.} Results show that asking for explanations of a specific question is one of the most frequent scenarios when referring physicians contact radiologists. As explainers, radiologists reported that they would like to ``\textit{know what the referring physician is looking for}'' and ``\textit{have a more focused question rather than a vague description}'' in order to generate more efficient and more specific explanations. Meanwhile, referring physicians looked for explanations for critical or atypical findings, \eg ``\textit{descriptions indicating high or low incidence of being significant}''.

\textit{With AI, physicians would seek explanations when their hypotheses are in discordance with AI's results}. Referring physicians reported more likely to need explanations when they believe they are right and AI's results are wrong or when they are not sure who is right.

\subsubsection{What Kinds of Explanation Are Needed}
\textit{Comparison with patients' prior images or other image modalities (\eg CT scan) is most commonly used for explanations,} as reported by both referring physicians and radiologists. Further, 18 participants considered more differential diagnoses as another way of explaining CXR-based results.

\textit{With AI, referring physicians further expect annotations on CXR and regional comparisons across patients.} 
Referring physicians requested more than what current AI's textual description of the observations, \eg ``\textit{... area of concern instead of ambiguous descriptions}''.

The needs for justifications emerged under this question. Different from explanation that explains with intrinsic process about how the results are generated, justifications apply extrinsic sources of information such as prior images or other image modalities to justify AI's results.
\section{Iteration \#2: Low-Fi Prototyping to Formulate Key System Designs}
Building off of the survey's findings that point to system requirements, we conducted a user-centered design to formulate CheXplain in two major iterations. In this section we describe the initial low-fi prototype co-designed with three physicians.

\subsection{Participants}
Recruited from a local medical center, the three physicians came from different specialties but all had experience reading radiographs. As shown in Table 1, the participants  had various degree of radiological knowledge, thus allowing us to calibrate system designs that serve for both novices and experts in radiology.

\begin{table}[h]
    \small
\centering
\label{survey participants}
\begin{tabular}{lcc}
\toprule[1pt] 
 \textbf{Participant}& \textbf{Specialty}  & \textbf{Radiological Knowledge} \\
 \midrule[0.5pt]
P1          & Pulmonologist &  Competent\\
P2                & Radiologist & Expert\\
P3                & Allergist & Novice\\
\bottomrule[1pt] 
\end{tabular}
\caption{Iteration \#2 physicians' knowledge in radiology. Novice: do not know much about radiology and do not report CXR images in workflow. Competent: know much about radiology but do not report CXR images in workflow. Expert: proficient in radiology and report CXR images in workflow.}
\end{table}

\subsection{Apparatus \& Procedure}
\begin{figure}[h]
    \centering
  \includegraphics[width=0.75\columnwidth]{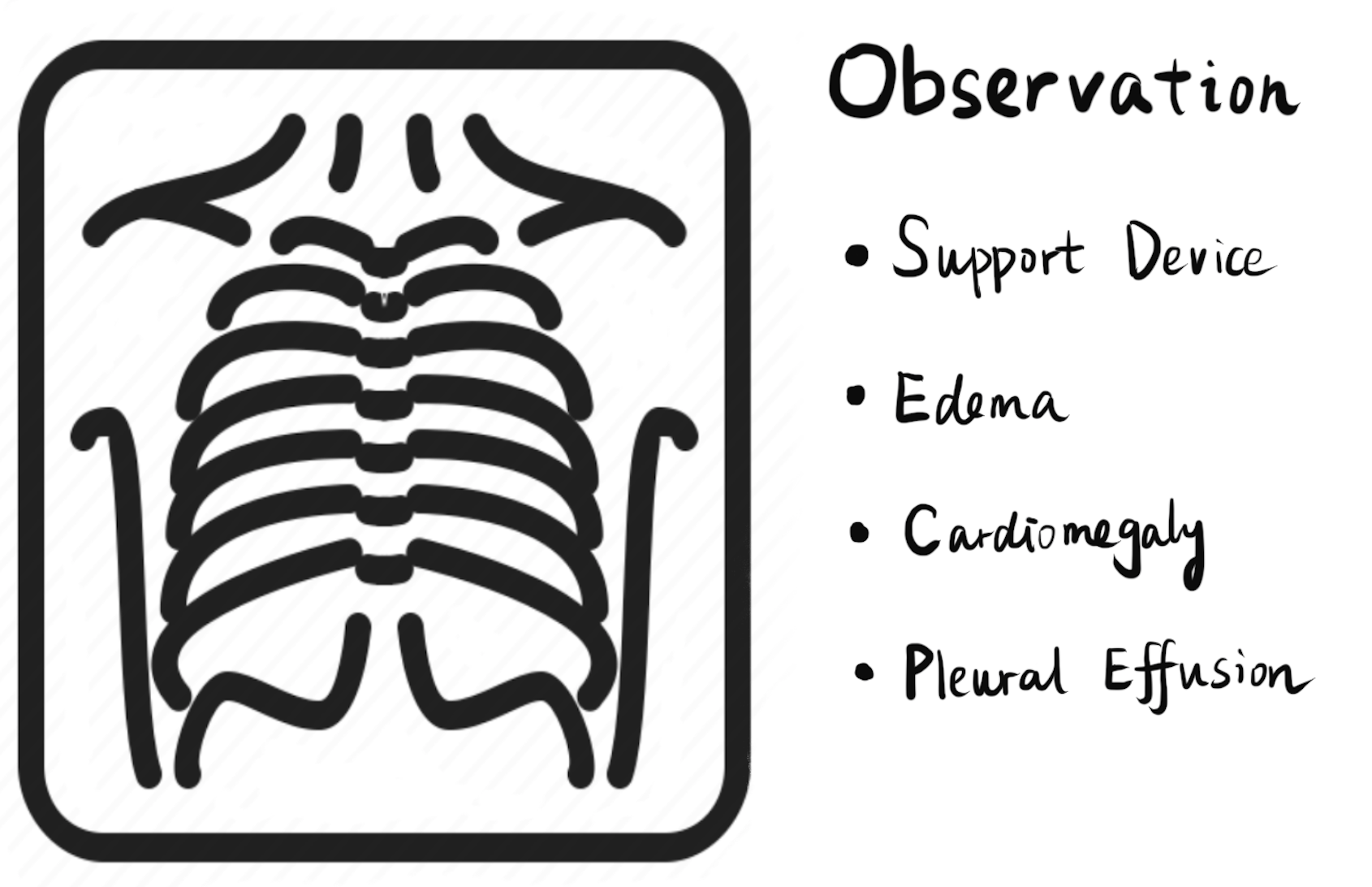}
  \caption{An example of AI-radiologist without any explanation to help participants understand the current status of AI radiologists}
  \label{CheXpert}
\end{figure}

To help participants understand the current status of `AI radiologists', we first presented a simple `system' based on the type of CheXpert's output without any explanation (Figure \ref{CheXpert}). Next, we used a combination of paper/pencil and Google Slides to create a series of low-fi prototype mock-ups. We used these artifacts as probes to elicit participants' reactions, critical feedback, and more discussions. Further, we asked participants to brainstorm more interactive features that could help them explore and understand an AI radiologist.

\subsection{Results: Key System Designs}
Through iterative development with three physicians, we consolidated the following key features of CheXplain. According to the findings in iteration \#1, feature \#1, \#2, \#8 are explanations and the others are justifications.

\begin{figure*}[h]
  \includegraphics[width=\linewidth]{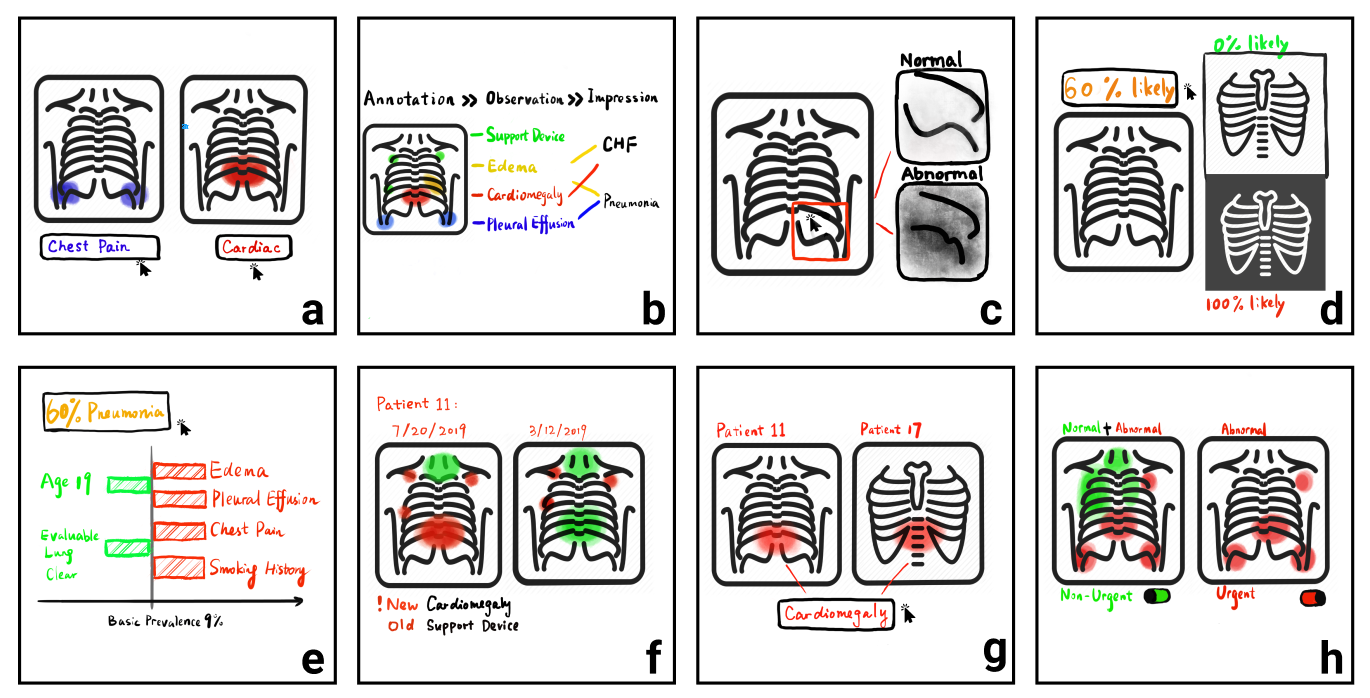}
  \caption{Low-fi prototypes of eight key system designs formulated in iteration \#2. 
  (\textit{a}) Augmenting CXR images with specific inquiries: directed explanation based on a user's target question; 
  (\textit{b}) Presenting explanations hierarchically: connecting low-level annotation, mid-level observation, and high-level impression; 
  (\textit{c}) Contextualizing observations with contrastive examples: contrasting a user-selected region with both normal and abnormal samples; 
  (\textit{d}) Interpreting observations probabilistically: calibrating the understanding of AI's probability by referring to same observations in other CXRs with different probabilities;
  (\textit{e}) Contextualizing impressions with prevalence \& traces: presenting prevalence and known factors that affect a user-selected impression;
  (\textit{f}) Comparison across time: showing the same patient's prior images;
  (\textit{g}) Comparison across patients: comparing with CXRs from different patients having the same observation;
  (\textit{h}) Mediating system complexity by urgency level: showing different amount of information in \textit{urgent} vs. \textit{non-urgent} mode.}

  \label{Low-fi}
\end{figure*}
\subsubsection{\#1 Augmenting CXR images with specific inquiries}
Prior AI-related work often considered a medical image as the only input and attempted to exhaust all possible diagnoses. However, as indicated in the survey and another paper \cite{bosmans2011radiology}, physicians often have a specific question for a CXR image, sometimes with important contextual information. P2 mentioned that ``\textit{For chest pain, they're usually looking for pneumothorax but for swallowing an object, they are looking for that object}''. P1 and P3 suggested that allowing for a range of questions/auxilliary information as additional input.

\insec{Design} As shown in Figure \ref{Low-fi}a, we design an initial step for referring physicians to select or type in keywords that characterize their questions specific to the patient's case. Accordingly, an AI should use such keywords to `filter' the examination process, \eg only paying attention to relevant regions of the CXR image.

\subsubsection{\#2 Presenting explanations hierarchically}
Previous work also mentioned that radiographical reports need to be itemized and better structured \cite{bosmans2011radiology}. According to our discussion with physicians and existing radiologists' guidelines, there are three levels of radiographical information: \one low-level examination---which body part to look at in a CXR image, \eg lungs, trachea; \two mid-level observation---what characteristics or phenomena are observed, \eg cardiomegaly, pleural effusion; \three high-level impression---differential diagnosis based on observations, \eg pleural effusion suggest pericardial disease or congestive heart failure.

Given such a hierarchy of information, we observed that more radiologically knowledgeable physicians tended to reason about AI's explanations in a bottom-up way while less knowledgeable physicians often took a top-down path when trying to make sense of AI.

\insec{Design} As shown in Figure \ref{Low-fi}b, the main interface of CheXplain embraces the three-level hierarchy: annotations on the CXR image indicate low-level examinations (where the AI is looking at); observations and impressions are presented next to one another. As a user selects a specific examination/impression/observation, the system `connects the dots': for example, by clicking \textit{Pneumonia}, the user can see that AI makes that impression based on \textit{Edema} observations, which are based on the examinations from CXR.

\subsubsection{\#3 Contextualizing observations with contrastive examples}
Physicians mentioned that comparing abnormal and normal CXR images is a common practice in teaching radiologists. Relatedly, XAI research has proposed contrastive \cite{dhurandhar2018explanations} and counterfactual \cite{hendricks2018generating} explanations that allow a person to understand something in light of plausible alternatives.

\insec{Design} As shown in Figure \ref{Low-fi}c, as a user selects a specific region of examination on the CXR image, the system shows a pair of contrastive examples also generated by the same AI examining the same region. By looking at images that contrast the AI's observation, a physician can explore `what-if' questions, \eg ``what if there is no cardiomegaly?'' By looking at images of the same observation, a physician can see if there is any common pattern, and whether the current image shares that pattern (if not, it might be a misdiagnosed outlier). 

\subsubsection{\#4 Interpreting observations probabilistically}
Many AI models can output a probability for each possible conclusion, which can be leveraged to explain the result. For example, a seemingly unlikely observation/impression is explainable if there is a low probability. However, physicians expressed confusion about interpreting a numeric probability, \eg ``\textit{Why AI is only 90\% sure, why is it not 100\% sure? I would like to see the gap between 90\% and 100\%}'' (P2). According to participants, we found that it was not the numeric values of probabilities that matter but how these values are mapped in medical workflows.

\insec{Design} As shown in Figure \ref{Low-fi}d, as a physician user clicks an observation label, the system shows an array of CXR images of the same observation, sorted by their probability and calibrated by medical professionals' understanding of an AI obervation's probability: unlikely (0\%), likely (50\%) and very likely (100\%). For example, a 80\% edema turns out to be quite similar to a near 100\% case.

\subsubsection{\#5 Contextualizing impressions with prevalence \& traces}
Participants brainstormed various ways to support the understanding of impressions: P1 suggested ``\textit{using epidemiologic standpoints}'' as a baseline to explain the probability of an impression; P2 addiontially suggested ``\textit{including the factors that AI took into consideration when generating the probability of impressions (e.g., the patient's age, risk factors, smoking history)}''.

\insec{Design} As shown in Figure \ref{Low-fi}e, the system provides two additional information to contextualize an impression: \one prevalence, \ie prior probability of the same impression in a given population; and \two traces that go back to observations that contribute to higher or lower probability of such impression.

\subsubsection{\#6 Comparison across time}
All three participants reacted positively to our initial idea of comparing with the same patient's prior images. P1 mentioned that an AI's observation is more explainable if the patient had similar diagnoses before. Similarly, P2 would like to see if the observations were already there in the prior images, and whether they were getting better or worse. P3 mentioned that previously non-existing observations should be prioritized when seeking explanations.

\insec{Design} As shown in Figure \ref{Low-fi}f, a physician user can switch to a side-by-side view between the current and prior cases, view annotations and filter to see only what is different.

\subsubsection{\#7 Comparison across patients}
To respond to the initial inquiry, participants suggested showing other patients whose cases had similar inquries.

\insec{Design} As shown in Figure \ref{Low-fi}g, in the same view as comparing with prior images, a physician user can see a lists of other patients' cases responding to cardiomegaly inquiry. 

\subsubsection{\#8 Mediating system complexity by urgency level}
Our survey shows that time cost contrains referring physicians from seeking explanations. While an explainable AI radiologist can mitigate such problem, sometimes physicians might not even have time to interact with our system. P3 mentioned, ``\textit{I will just pick the important abnormalities if I am in a hurry and look back to see other information later}''. P1 suggested two different modes of the system for urgent and non-urgent cases.

\insec{Design} As shown in Figure \ref{Low-fi}h, we design an `urgent' mode that can be toggled by the user, where the system only shows annotations to explain the most significant AI-generated observations, \ie observations that have both high confidence {\it and} lead to impressions of critical nature.

\section{Iteration \#3: Evaluating a High-Fi Prototype}
We integrated the eight key features into a high-fidelity prototype of CheXplain, which allowed us to further iterate the design by having physicians interact with the system to explore and understand the AI-generated radiograph analysis.

\subsection{Physicians}
We invited 39 referring physicians who participated in the survey study and other medical professionals via connections to a local medical center. We finally recruited six participants. As shown in Table 2, the participants had various degree of radiological knowledge.
\begin{table}[h]
    \small
\centering
\label{high-fi participants}
\begin{tabular}{lcc}
\toprule[1pt] 
 \textbf{Participant}& \textbf{Specialty}  & \textbf{Radiological Knowledge} \\
 \midrule[0.5pt]
P1          & Pediatrics &  Competent\\
P2                & Internist & Expert\\
P3                & Pediatrics & Competent\\
P4                & Medical student & Novice\\
P5                & Pediatrics & Competent\\
P6                & Medical student & Novice\\

\bottomrule[1pt] 
\end{tabular}
\caption{Iteration \#3 Participants' knowledge in radiology. Novice: do not know much about radiology and do not report CXR images in workflow. Competent: know much about radiology but do not report CXR images in workflow. Expert: proficient in radiology and report CXR images in workflow.}
\end{table}


\subsection{Apparatus}

\begin{figure}
  \includegraphics[width=\linewidth]{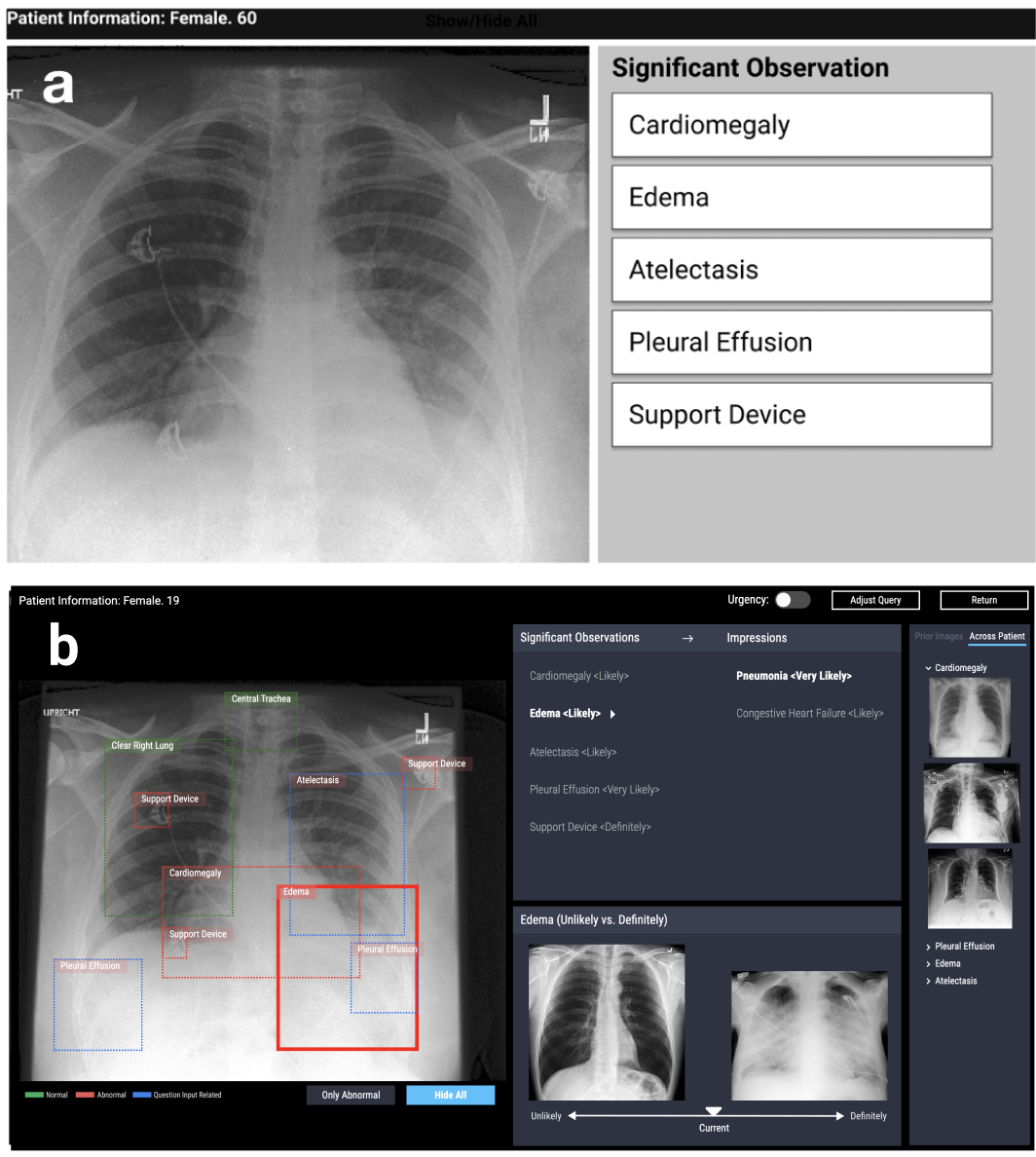}
  \caption{High-fi prototyping. (\textit{a}) CheXpert: AI radiologist without explanations. (\textit{b}) CheXplain: The high-fidelity prototype with 8 key features.}
  \label{CheXplain}
\end{figure}

Figure \ref{CheXplain} shows the study testbed system. We implemented the front-end of CheXplain as a web-based platform. All patient cases including input, regional annotations, prior and across-patient images were from the CheXpert dataset. Similarly, CXR observations were generated by a convolutional neural network used in CheXpert paper. The remaining requisite information was manually generated based on external medical guidelines \cite{editionunofficial} and was verified by a radiologist collaborator. Using manually-generated data to test a fully-implemented front-end allowed us to agily test the system design without committing to a computationally costly implementation of the back-end.

Physicians interacted with CheXplain using their own browsers. All the studies were conducted remotely via Zoom through which we observed physicians' interactions in real-time via screen sharing. We recorded the sessions with physicians' permission and transcribed the audio for further data analysis. 

\subsection{Tasks \& Procedure}
We started with a brief introduction of AI's application in medical image analysis. Next, physicians were presented with a patient's case randomly selected from the database, which came with textual labels indicating AI-generated observations. The main task was to interpret AI's results, \ie describing their understanding of why AI arrived at certain results of the case. We asked the physicians to respond as succintly and accurately as possible. 

We asked each physician to describe their interpretation of AI after presenting CheXpert's results (Figure \ref{CheXplain}a). Then, we guided the physician to interact with CheXplain (Figure \ref{CheXplain}b), using it as a tool to explore and understand AI-generated results of the same patient's case. Afterwards, we asked the physician to describe their interpretation of AI again.

Physicians were also encouraged to think aloud during their interaction with CheXplain. We asked physicians {\it in situ} questions as they interacted with each of the eight key features---whether and how each feature helped them understand AI's results. We ended by asking each physician to summarize how CheXplain overall changed their understanding of the underlying AI, and whether and how such systems can be integrated into their existing workflow.

\subsection{Analysis \& Findings}
We analyzed each physician's response at a rolling basis. The first experimenter transcribed and coded physicians' responses and iteratively developed a code book. A second experimenter reviewed the codes and resolved disagreements via discussion with the first experimenter.

\subsubsection{How physicians interpreted AI before \& after CheXplain}
We compare physicians' interpretation of AI before and after interacting with CheXplain. Results show that pre-CheXplain responses (\ie only relying on CheXpert's results) contain very little information---all but one physicians acknowledged that they were not able to provide an interpretation. The only physician who did respond described that ``\textit{AI gathered a lot of images with readings similar to human radiologists}'', which was based on his previous knowledge of AI. In contrast, responses contained more substance after interacting with CheXplain. All physicians mentioned that AI arrived at the observations based on comparison against normal/abnormal images and prior images, which captured the data-driven, learning-based nature of AI. In addition, three physicians also mentioned that AI pointed out where it was looking at (P3, P5, P6)---``\textit{It would give you a better idea because it tells you what areas to look at if you are not sure about the analysis and this is exactly what I would be looking for}'' (P4). 
\begin{figure}[h]
  \includegraphics[width=\linewidth]{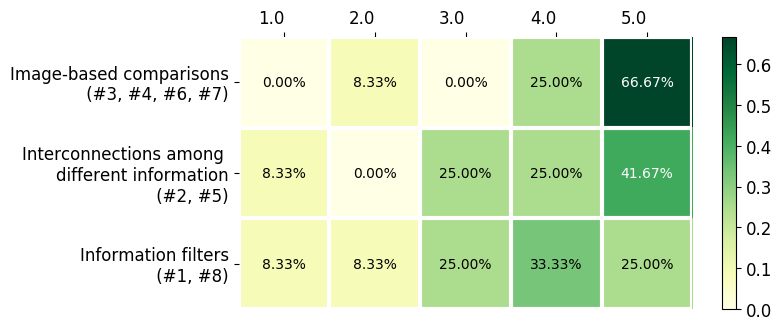}
  \caption{Percentages of physicians' ratings for grouped features. The vertical axis presents groups in a top-down descending order by mean values of ratings.}
  \label{rating}
\end{figure}

\subsubsection{Overall feedback: how CheXplain enables understanding AI}
Participants were asked whether and how each feature helped them understand AI's result as well as rating the helpfulness from 1 (not helpful at all) to 5 (extremely helpful). We group the aforementioned eight design features into three different categories: image-based comparisons, interconnection among different information, and information filters. As shown in Figure \ref{rating}, image-based comparisons were regarded as the most helpful features. Below we provide detailed qualitative discussion about each of the three categories
(numbers in parentheses refer to individual features discussed in Iteration \#2).
\subsubsection{Image-based comparisons (\#3, \#4, \#6, \#7)}

Image-based comparisons help physicians to understand AI by drawing on specific CXR examples as evidence, which helped physicians overcome the difficulty of understanding AI's results in isolation. Though image comparisons are common in image studies, CheXplain enables physicians to screening out whether AI makes mistakes in each step by comparisons.
For example, \textit{Interpreting observations probabilistically} by calibrating physicians' understanding of probability is helpful because offering images with different probabilities ``\textit{defines different likelihoods so I know which is statistically more likely}'' (P1) while preventing users from ``\textit{missing something that's really obvious in AI's definitive category}'' (P2). 
\textit{Contextualizing observations with contrastive examples} was helpful by ``\textit{constrasting regions, I can tell what a normal region looks like, determine how similar the images are and understand AI's conclusion}'' (P5). 
\textit{Comparison across time} pointed out the difference of AI's observations from prior images, which might seem unlikely if viewed in isolation (P3, P6) and \textit{Comparison across patients} surfaced common features for the same observation across multiple patients (P1, P2, P6). 

However, there were confusions when physicians tried to understand AI-based on multiple sources of images. P1 and P2 thought AI also analyzed the patient's prior images when processing the present case, which suggests that design should clearly show what is and is not the input data to the AI.
P3 mentioned that a common feature of the same observation might not always exist because sometimes it appears very differently across patients. To address this, future design should group cross-patient cases based on their variance so that physicians can look for common features only in the same group.

\subsubsection{Interconnection among different information (\#2, \#5)}
Although establishing interconnection of information is not novel in medical systems \cite{gritzalis2004security}, we initially employ it in medical explainable AI. Participants found it useful for CheXplain to present AI's findings in an interconnected way, which serves as a road map for them to navigate CXR results at different levels of detail, although not all the interconnected information was found equally helpful for their understanding of AI. 

For example, \textit{presenting explanations hierarchically} provided an organized way for physicians to interpret AI's output, from which low-level annotations were found to be more helpful---``\textit{I can know where AI looked at to figure out the result and it tells you exactly what's there, where, and why}'' (P6). Impressions, on the other hand, were more controversial and often considered less helpful: P4 believed generating impressions is more complicated than what the system shows, which simply traces an impression to the lower-level observations and examinations. \textit{Contextualizing impressions with prevalence \& traces} was considered beyond the scope of the system. While physicians liked the concept of having such contextual information for each impression, most thought it was ambitious to explain both radiological and clinical factors, and were concerned that other important factors might be missing here. Future design can adopt an on-demand approach, \eg allowing physicians to ask specific questions about what they believe are relevant to contextualize a given impression.

\subsubsection{Information filters (\#1, \#8)}
Physicians found that both features in this category were not directly related to helping physicians understand AI; rather, physicians considered them as filters to help physicians target their understanding or reduce time cost of interacting with the system. Despite almost all AI research in medical domain claiming that they speed up the diagnosis, which are actually provided by computing power, we start from a physician-centered perspective and improve this problem by interaction. 

Physicians mentioned that \textit{Augmenting CXR images with specific inquiries} and \textit{Mediating system complexity by urgency level} helped narrowing down the information when understanding AI and sorting significant observations fast---''\textit{they help me to focus on observations that AI thinks are more likely to be related to my input questions}'' (P5). P3 noted that AI presents findings related to his question input in the image.

\textit{Mediating system complexity by urgency level} opened up physicians' various preferences in filtering information: P4 would like to ``\textit{see the important observations faster}'' when running out of time; P6, on the other hand, ``\textit{still wants to see everything to get a better understanding}''; P3 preferred to ``\textit{pick what I think is urgent}'' (P3). Further design could address these diverse requirements by making this feature customizable based on physicians' personal preferences. 

\section{Design Recommendations for Medical AI System}
Based on our survey and a user-centered process of prototyping CheXplain, we summarize design recommendations for developing medical AI systems that can be understandable to physicians. Specifically, there are four recurring themes across our three iterations of design activities: motivation vs. constraint and explanation vs. justification.

\subsection{Motivation vs. Constraint}
Motivation is what drives a physicians to try to understand an AI's diagnosis results, and constraint is what prevents them.

{\bf Enable a physician to achieve an understanding of AI targeted at a specific clinical problem}, rather than to attain a general, open-ended understanding of AI. 
Our survey indicates that a lack of information motivates physicians to seek explanations; however, when co-designing CheXplain, physicians tended to focus on understanding aspects of AI's results only if they are relevant to their own hypothesis or differential diagnosis. Besides, in radiographic reporting, referring physicians also need to state clearly what clinical question they want to have answers to \cite{bosmans2011radiology}. 
To address this, we recommend starting with a question input to narrow the scope of AI's analysis. Such a question-answering (QA) ability also presents implementation challenge to the medical AI community given the recent development of visual QA \cite{antol2015vqa,park2016attentive,selvaraju2016grad}. 
Currently, one potential pitfall is attempting to explain too much, or too deep into the technical stack, which might actually constrain a physician's ability to see and understand the more clinically relevant aspects of AI.

{\bf Make a wide range of data sources available as a physcian tries to understand the AI}, as differential diagnosis seldom relies on a single type of data \cite{Xie2019-qf}. We recommend going beyond how a medical AI currently handles one specific type of input modality only (\eg CheXpert only looks at CXR images). A mixed-modality approach is more familiar to physicians and more likely to motivate them to understand AI's results. 
In contrast, one potential pitfall is developing multiple XAI systems for multiple types of data, \eg System A explains mining eletronic health records while System B explains deep convolutional network' processing of MRI data, which is likely to constrain a physician's effort in understanding how a clinical case is handled by AI.

{\bf Let physicians customize how much time they want to spend trying to understand AI}. CheXplain provides an urgency toggle for physicians with low time budget to understand AI; yet this design still cannot address different approaches physicians take to juggle time and information. We recommend that, instead of adjusting time, the system should let a physician freely control the amount of information they would like to process, from trying to make sense of only the critical results, to revealing more explanation hierarchically, and to referencing other regional, previous or cross-patient examples.


\subsection{Explanation vs. Justification}
Our research started with a focus on explanations of AI, which enables physicians to understand AI with {\it intrinsic} processes of producing certain output given certain input data, \eg processing pixels of a CXR image to arrive at certain observations. Later, as we co-designed CheXplain with physicians, a different set of ideas emerged, which tried to draw on {\it extrinsic} sources of information to justify AI's results. Such justifications include prevalence of a disease, contrastive examples from other CXR images, and a patient history. Thus, we consider explanations and justifications as two categories of methods that help physicians understand AI.

{\bf Use more explanation for medical data analysts (\eg radiologists), and more justification for medical data consumers (\eg referring physicians)}. We find that radiologists tend to expect more explanations for details such as low-level annotations of CXR images, while referring physicians are generally less concerned about the instrinsic details but care more about extrinsic validity of AI's results. Retrospectively, we can see that five of CheXplain's features are justification (\#3-7).

{\bf Enable explanation and justification at different levels of abstraction}, similar to how CheXplain employs the examination-observation-impression hierarchy to scaffold both explanation and justification. Holistically, as a physician follows a bottom-up or top-down path to understand an AI's diagnosis of a patient, at any step along the way, they should be able to seek both explanation and justification. To achieve this, the XAI community need to consider explanation regulated by a user-specified level of abstraction; research on Content-Based Image Retrieval (CBIR) should enable search criteria at multiple levels of abstraction, \eg from a region of specific visual features to a global image that presents a specific impression.



{\bf Implementation challenges for medical imaging AI: explain how AI `looks at' a region; justify how two images are clinically (dis)similar}. 
    
    A plethora of prior work \cite{dabkowski2017real,fong2017interpretable,selvaraju2016grad,simonyan2013deep} has achieved localizing where AI is `looking at' on an input image, but not {\it how}---\eg how does AI consider this region as edema? Generating a natural language explanation---analogous to Hendricks' work on bird classification \cite{hendricks2016generating}---presents a new challenge for medical imaging AI. 
    
    When comparing images to justify AI's result, one challenge is to inform physicians how two images are radiographically (dis)similar, thus further enable them to gain insight from such justifications. Cai \etal designed a TCAV-like \cite{cai2019human} tool for pathologists to find images based on self-defined concepts, which should be generalized to enable justifications beyond comparing visual features.
\section{limitation}
Our current work has the following limitations:
\begin{itemize}
    \item Survey questions regarding an AI radiologist were inevitably speculative due to the limited adoption of AI in current medical practices;
    \item The number of participants was limited due to the scarcity of medical professionals' availability;
    \item The available CXR dataset we used has limited information to further the exploration using CheXplain, \eg missing patients' medical history, no additional lab results.
    \item The final evaluation of system design is inevitable speculative instead of in a real workplace due to strict regulations. The results may have slight differences in the actual diagnostic workflow.
\end{itemize}
\section{Conclusion}

In this paper, we identified the gap between current XAI research and domain experts' understanding of AI in medical decision-making. Our work contributes to a bridge of this gap by conducting a physician-centered design of CheXplain --- a system that enables physicians to understand AI-enabled chest X-ray analysis by interaction. Ultimately, we propose a series of recommendations for future human-centered medical AI development around four recurring themes: motivation vs. constraint and explanation vs. justification. 

Taken together, our work provides implications for how physicians can explore and understand data-driven, AI-enabled medical imaging analysis to assist physicians in medical decision-making process. The future work will explore the development of human-centered medical AI beyond medical imaging analysis.
\section{Acknowledgement}

This work was funded in part by the National Science Foundation under grant IIS-1850183. 

We thank Fereidoun Abtin and Jonathan Barclay for their valuable comments on earlier design of CheXplain. We thank all the anonymous participants for their contributions to our study. We also thank Lauren Hung for her suggestions on interface design of the system.

\balance{}

\bibliographystyle{SIGCHI-Reference-Format}
\bibliography{references}

\end{document}